# Percolation Transition in the Heterogeneous Vortex State in NbSe$_2$

O. Dogru and E. Y. Andrei

Department of Physics and Astronomy, Rutgers University, Piscataway, New Jersey 08855

M. J. Higgins

NEC Research Institute, 4 Independence Way, Princeton, New Jersey 08540

S. Bhattacharya

Tata Institute of Fundamental Research, Mumbai 400-005, India

A percolation transition in the vortex state of a superconducting 2H-NbSe$_2$ crystal is observed in the regime where vortices form a heterogeneous phase consisting of ordered and disordered domains. The transition is signaled by a sharp increase in critical current that occurs when the volume fraction of disordered domains, obtained from pulsed measurements of the current-voltage characteristics, reaches the value $P_c = 0.26 \pm 0.04$. Measurements on different vortex states show that while the temperature of the transition depends on history and measurement speed, the value of $P_c$ and the critical exponent characterizing the approach to it, $r = 1.97 \pm 0.66$, are universal.

Recent imaging experiments[1] on vortex matter in superconductors[2] have revealed a heterogeneous state consisting of domains of ordered and disordered phases[3,4,5]. The appearance of heterogeneity coincides with the observation of striking anomalies in transport and magnetization measurements including the peak effect[6], nonlinear response, metastability, generation of broadband noise and history effects[7,8,9]. It is well known that multi component systems undergo a percolation transition when the volume fraction of one of the components reaches a critical value $P_c$ leading to the formation of a system-spanning cluster and to critical behavior[10]. In the case of the heterogeneous vortex system, because the critical current depends on the degree of order, the transport properties are sensitive to details of the spatial domain distribution and can lead to dramatic changes in the properties of the superconducting host when one of the domains percolates.

In this letter we report on experiments demonstrating the existence of a percolation transition in the vortex system in $NbSe_2$ and its effect on the transport properties. The experiments employed pulsed measurements of the voltage-current (*V-I*) characteristics together with a proposed two-phase model to obtain the fraction of disordered domains *P*. We find that the transition, is signaled by a sharp rise in $I_c$, that occurs at $P_c = 0.26 \pm 0.04$ and that the approach to $P_c$ is characterized by a critical exponent $r = 1.97 \pm 0.66$. Our experiments show that the transition is uniquely determined by $P_c$ and not by the thermodynamic parameters. Thus while the transition temperature varies with magnetic field and can be lowered with the application of a slow current ramp or increased by applying an ac current[11,12] the values of $P_c$ and *r* are universal.

We measured two *Fe*-doped *2H-NbSe$_2$* crystals. Sample A with dimensions 3x0.7x0.03 mm$^3$ critical temperature $T_c$ = 6.03 K, and transition width $\Delta T_c$ = 50 mK and sample B with 3x1x0.015 mm$^3$, $T_c$ = 5.61 K, $\Delta T_c$ = 40 mK. All the data presented here was taken on sample A. The data for sample B, although less extensive, was in good agreement with that for sample A. The field was applied in the *c* direction, perpendicular to the sample surface, and the current was in the *ab* plane. The experiments employed a standard 4-lead configuration with *AgIn* solder contacts to monitor the voltage response to current ramps and pulses. A low noise (4nV/ Hz$^{1/2}$) fast amplifier was used to detect the voltage signal which was then digitized with a 100 MHz digital oscilloscope. The response time of the entire system including leads and sample in the normal state was 2 µs. In order to reduce noise the data were averaged over 10 runs (initiating the system with a zero field cooling cycle from above $T_c$ for each run). A 2 µV voltage response criterion was used to define critical currents. The current-voltage (*V-I*) characteristics were obtained point by point from the time evolution of the voltage response to a current pulse as illustrated in Fig. 1. Each point on the *V-I* represents the voltage measured 3 µs after applying the current pulse. This procedure excludes the instrumental response which decays within ~2 µs of the change in current amplitude. Repeating the measurement for another current value gives an additional point in the *V-I* curve and so on. Each point is obtained on a pristine vortex lattice freshly prepared from the normal state. It was previously shown that for samples and mounting methods identical to the ones used here Joule heating effects on time scales < 10 µs are negligible[13]. In addition to avoiding heating, the pulsed technique also eliminates effects due to current induced organization as we show below.

An applied current can induce reorganization of the vortex lattice by transforming one phase into another[7,14] or introducing a disordered phase as new vortices penetrate through the surface barrier at the sample edge[15]. Thus, unless the measurement is faster than the organization process, the result depends on the measurement speed. The relevant time scale for current induced reorganization can be estimated from $\tau \sim a/v$, the time to move a lattice spacing $a$. Here $v = V/Bl$ is the vortex velocity and $l = 2$ mm the distance between voltage leads. For $B = 0.5$ Tesla and $V = 2\mu V$ corresponding to the voltage resolution, we find $\tau \sim 10 \mu$s which is longer than the measurement time in the pulsed experiments. By comparison in a current ramped measurement $I_c$ is reached within a time $\tau = I_c/R$, where $R$ is the ramping rate. For example when $R = 1$ mA/s (henceforth labeled as Slow Current Ramp- SCR) the time to reach a typical value of $I_c = 50$mA, $\tau \sim 50$ s, is much longer than the reorganization time. For our Fast Current Ramps (FCR) where $R = 300$ A/s we find $\tau \sim 150$ μs which still allows for reorganization.

In order to calculate $P$ in the heterogeneous state we will need the values of $I_c$, for ordered and disordered states. The ordered lattice was obtained at low temperatures with Zero Field Cooling (ZFC) where the field is applied after cooling through the superconducting transition. Far below the peak effect the ZFC state is ordered, stable, and $I_c$ is independent of measurement speed[7,11,14] as shown in Fig. 2 where we compare the temperature dependence of $I_c$ obtained with various measurement speeds. At low temperatures we find that $I_c$ for the ZFC state is the same for FCR, SCR (solid and open triangles) and pulsed (not shown) measurements. The disordered lattice can be prepared by Field Cooling (FC), where cooling from the normal state is carried out in the presence of a magnetic field[7]. For the doped sample used here the FC state is a robust supercooled disordered state that does not evolve with time due to the weakness of thermal fluctuations compared to the pinning potential[7 9 11]. But because it is metastable, the FC state is easily driven by a current to reorganize, leading to a strong dependence of $I_c$ on measurement speed as illustrated in Fig. 2. As can be seen in the inset of Fig. 2 the voltage response does not evolve much immediately following the 2 μs duration of instrumental transients (the same transient with opposite sign appears when the current is removed as shown in the inset of Fig. 1), but it grows significantly for longer times. We therefore assume that the voltage recorded 3 μs after the current onset represents the response of the pristine FC lattice before it has reorganized. In the analysis described below we will assume that the low temperature values of $I_c$ (at $T_0 = 4.3$ K) for the ZFC and the pulsed FC represent the critical currents of ordered and disordered states respectively.

We now describe the protocol to measure $P$. The sample was prepared by ZFC at a (variable) target temperature $T$. Subsequently, after waiting for ~1 minute at $T$, and without changing the field, the sample is rapidly cooled (over several seconds) to $T_0 = 4.3$ K where the $V$-$I$ characteristics was measured point by point with the pulsed technique. The pulsed $V$-$I$ at 0.5 Tesla and several values of $T$ are presented in Fig 1b. We note that the slope of each curve increases monotonically saturating at a value of ~2.4 mΩ, which corresponds to the expected free flux flow resistance at $T_0$. We note that although the measurement temperature was the same the shapes of the $V$-$I$ depend on the preparation temperature $T$. We will show that the shape of these curves together with the values of $I_c$ in the ordered and disordered states gives the value of $P(T)$.

In the two-phase model the domains of Ordered Phase (OP) and Disordered Phase (DP) are characterized by critical current densities $J_o$ and $J_d$ respectively with $J_d > J_o$[16]. Before vortices start moving the current density at any given position inside the sample has to be either zero or equal to the local critical current density, which can take one of the two values $J_d$ or $J_o$. Vortex motion first occurs at an applied current for which there exists a cross section (transverse to the current at every point) within which all vortices are subject to their respective critical current density. If the cross section $A(x)$ at position $x$ along the current flow contains a fraction $\alpha(x)$ of DP, the critical current for that cross section is

1. $I_c(x) = (\alpha(x)J_d + (1-\alpha(x))J_o))A + I_s$ .

Here $I_s$ is a surface current resulting from the Bean-Livingston surface barrier[15][17]. We have assumed that both the cross section and the surface current are uniform along the sample so that $A(x)=A$ and $I_s(x) = I_s$ , but the model can be generalized to a situation where this is not so, including the case of 3D vortex lattices, as long as vortices do not intersect. The global critical current, obtained by minimizing $I_c(x)$ over all cross sections, is then determined by the cross section containing the minimal fraction of DP, $\alpha_m$. As long as a cross section exists that does not intersect a domain of DP, $ie$ $\alpha_m = 0$, the critical current will be the same as that of a sample with all vortices in the ordered phase $I_o = J_oA + I_s$. Just above the percolation transition, after the first system-spanning cluster of DP has formed, cutting through all cross sections, $\alpha_m > 0$ and $I_c > I_o$.

Thus we can obtain the value of $\alpha_m$ for an arbitrary distribution of domains from the values of $I_c$, $J_o$ and $J_d$. Although the critical currents $I_o$, $I_d$ of the ordered and states respectively can be measured as discussed above this is not the case for the critical current *density* $J_c = (I_c - I_s)/A$ which contains a surface component not directly measured. The presence of $I_s$ can lead to a significant over-estimate of $J_c$, especially in low pinning superconductors, if one calculates it by assuming a homogeneous current distribution. Since direct measurements of $I_s$ are quite difficult it is useful to work with quantities that do not explicitly depend on it. We note that $I_s$ is the current needed to overcome the attraction between a vortex and its virtual image in the boundary and therefore it is a single-vortex property which should not depend on the degree of order in the vortex state. It is therefore reasonable to assume that $I_s$ would be the same for the OP, DP and the heterogeneous state. This would then lead to a cancellation of $I_s$ from equation 1 resulting in a simplified expression for $\alpha_m$ which involves only the measured quantities at one temperature: $I_c(T) = I_o + \alpha_m(I_d(T) - I_o(T))$. To take advantage of this simplification all measurements were carried out at the same temperature $T_0 = 4.3$ K. Heterogeneous states with different compositions of DP can be obtained by varying the temperature at which the ZFC states are prepared. This becomes clear by noting that the *V-I* curves in Fig. 1b are qualitatively different from each other even though they are measured at the same temperature. In other words preparing ZFC states at $T$ and then cooling to $T_0$ creates a frozen replica of the state at $T$. This is consistent with earlier studies on similar samples that showed that the FC procedure leads to a robust supercooled vortex state[7,11]. It is also consistent with the fact that the two phases must be separated by an energy barrier that far exceeds the thermal energy in order to exhibit the experimentally observed stability and coexistence[1]. The assumption that the vortex lattice does not change its composition upon further cooling implies $\alpha_m(T_0) = \alpha_m(T)$ and therefore $\alpha_m(T)$ can be obtained directly from the measured critical currents at $T_0$:

2. $\alpha_m(T) = (I_d(T_0) - I_o(T_0))/(I_c(T,T_0) - I_0(T_0))$.

Here $I_c(T,T_0)$ is the critical current of the ZFC state prepared at $T$ and measured at $T_0$ and $I_o(T_0)$, $I_d(T_0)$ are the critical currents in the OP and DP measured and prepared at $T_0$. To test the validity of the model we used this value of $\alpha_m$ to calculate the critical current of the ZFC state at $T$: $I_c(T) = I_o(T) + \alpha_m (I_d(T) - I_o(T))$ and compared it to the directly measured value. The results of this calculation (open circles in Fig.2) are in good agreement with the measured values of $I_c(T)$ (solid triangles in Fig. 2), confirming the assumption that the composition of the vortex lattice does not change upon cooling to $T_0$.

In order to obtain the value of $P$ it is necessary to go beyond the critical current and analyze the shape of the $V$-$I$ curves. In our model the voltage response to an applied current $I$ can be expressed as $V(I) = R_{ff} \int_0^1 (I - I_c(x)) dx$ where $x$ is in units of sample length and the integrand is zero for $I < I_c(x)$. By using Equation 2 and the fact that $P = \int_0^1 \alpha(x) dx$, we obtain the voltage response in the free flux flow regime where all vortices are in motion $V(I) = R_{ff}(I - I_o - (I_d - I_o)P)$. Solving for $P$, we obtain

$P = \dfrac{I_{eff} - I_o}{I_d - I_o}$ where $I_{eff}$ is the $V = 0$ intercept of the linear portion of the $V$-$I$ curve. This gives a straightforward geometrical procedure to obtain $P$ from the $V$-$I$ curve.

Alternatively $P$ can be obtained from the curvature of the V-I curves

$$P = \frac{1}{R_{ff}(I_d - I_o)} \int_{I_c}^{I_{cf}} (I - I_0) \frac{\partial^2 V}{\partial I^2} dI.$$ Here, $I_{cf}$ is the current at which the slope recovers the free flux flow resistance. Both methods give the same values for $P$ within experimental accuracy.

In Fig. 3a we plot the temperature dependence of $P$. We note that for $P < 0.27$, they are equal indicating that the disordered phase has not yet formed a system-size cluster. The percolation threshold as signaled by the onset of increasing $I_c$, occurs at $P_c = 0.27 \pm 0.04$. Beyond this point $I_c$ continues to grow, as the disordered regions expand with increasing $P$, until the entire sample becomes disordered. It is interesting to note that, although the vortex states prepared at $T = 4.98, 4.992$ and $5.006\ K$ have the same $I_c$, the shape of the V-I curves shown in Fig. 1b are quite different reflecting the different contents of DP.

In Figure 4a we plot $\alpha_m$ calculated from Eqn. 1 as a function of $P$ together with a power law fit to $\alpha_m \propto (P - P_c)^r$. The exponent $r$ characterizes the behavior of the minimal cross section close to $P_c$ and is not one of the standard critical exponents. We can however show that the behavior of $\alpha_m$ near $P_c$ can be mapped onto the conductivity near the metal-insulator transition and therefore $r$ should be the same as the conductivity exponent $t$[6,18]. The analogy is drawn by considering a binary insulator-conductor sample that is topologically identical to the heterogeneous vortex sample. This is obtained by mapping all the DP domains onto a conducting phase with conductivity $\sigma_c$ and all the OP ones onto an insulating phase. Just above the percolation transition the conductivity is dominated by the bottlenecks at the cross section containing a minimal amount of conducting phase, $\alpha_m$, so that the effective conductivity is $\sigma \propto \sigma_c \alpha_m$. This is valid in the limit where the bottlenecks are much narrower than any other conducting cross section. Near the percolation transition the conductivity will thus be $\sigma \propto \alpha_m \propto (P - P_c)^t$ where $t$ is the conductivity exponent. In 3D systems the value $t = 2$ is expected to be universal[6]. Fitting our data over a range of $P$ close to $P_c$ we find $r = 1.78 \pm 0.52$ which is consistent with our model.

A salient feature of percolation is that it is governed by a single parameter $P_c$ and thus provides a stringent test for identifying the transition. To check the validity of the vortex percolation interpretation we repeated the experiment with vortex states prepared by different methods. Current annealed vortex states were prepared by applying a slow current ramp to the ZFC state at $T$ before cooling to $T_0$. The onset of increase in $I_c$ for the annealed states (open triangles in Fig 2) is shifted to a lower temperature compared to the un-annealed ZFC states, yet if this is the signature of percolation we should find the same value of $P_c$. The results obtained for the annealed lattices are presented in Figure 4b. Although the temperature of the transition 4.7 K[19] is lower than that of states prepared without annealing, the percolation threshold $P_c = 0.24 \pm 0.02$ and the exponent $r = 2.01 \pm 0.32$ are in good agreement with the previous values. Another set of measurements carried out on the un-annealed ZFC vortex lattice at B = 0.75 T (Figure 4c) resulted in $P_c = 0.27 \pm 0.02$ and $r = 2.14 \pm 0.25$. Again these results are in good agreement with the data at B = 0.5 Tesla confirming the universal nature of the transition[20]. Comparing to other systems we find that $P_c \sim 0.26 \pm 0.04$, the value of the percolation threshold averaged over the three experiments, places the vortex percolation transition in the universality class of overlapping random spheres (the "inverse Swiss cheese" model)[6].

The experiments described here demonstrate that the onset of increase in critical current associated with the peak effect is the signature of a percolation transition when domains of disordered phase form a system-spanning cluster. The percolation transition is uniquely determined by the volume fraction of disordered phase, $P_c$, and unlike the onset temperature of the peak effect, it does not depend on history, measurement speed or method of preparation. Because the value of $P_c$ at the transition is universal it provides an unambiguous description of the onset of the peak effect.

We wish to thank Z.L. Xiao, G. Li, N. Andrei, H. Kojima and I. Skachko for stimulating discussions. This work was supported by DOE grant DE-FG02-99ER45742
.

---

[1] M. Marchevsky, M. J. Higgins, and S. Bhattacharya, Nature (London) **409**, 591 (2001); A. Soibel et al. Nature (London) **282**, 406 (2000).

[2] G. Blatter et al. Rev. Mod. Phys. **66**, 1125 (1994); E. H. Brandt, Rep. Prog. Phys. **58**, 1465 (1995); T. Nattermann and S. Scheidl, Adv. Phys. **49**, 607 (2000).

[3] P. L. Gammel, L. F. Schneemeyer, J. V. Waszczak, and D. J. Bishop, Phys. Rev. Lett. **61**, 1666 (1988); E. M. Forgan, Nature (London) **343**, 735 (1990); R. Cubitt, Nature (London) **365**, 07 (1993).

[4] D. R. Nelson, Phys. Rev. Lett. **60**, 1973 (1988); M. C. Marchetti and D. R. Nelson, Phys. Rev. B **41**, 1910 (1990); T. Giamarchi and P. LeDoussal, Phys. Rev. B **52**, 1242 (1995); D. Li and B. Rosenstein, Phys. Rev. B **65**, 220504 (2002).

[5] Y. Paltiel et al. Nature (London) **403**, 398 (2000).

[6] S. H. Autler, E.S. Rosenblum, and K. H. Gooen, Phys. Rev. Lett. **9**, 489 (1962); W. DeSorbo, Rev. Mod. Phys. **36**, 90 (1964), M. J. Higgins and S. Bhattacharya, Physica C, **257**, 232 (1996).

[7] W. Henderson, E. Y. Andrei, M. J. Higgins, and S. Bhattacharya, Phys. Rev. Lett. **77**, 2077 (1996).

[8] A. C. Marley, M. J. Higgins, and S. Bhattacharya, Phys. Rev. Lett. **74**, 3029 (1995); G. D`Anna et al., Phys. Rev. Lett. **75**, 3521 (1995).

[9] R. Wordenweber, P. H. Kes, and C. C. Tsuei, Phys. Rev. B **33**, 3172 (1986); S. Bhattacharya and M. J. Higgins, Phys. Rev. B **52**, 64 (1995); S. S. Banerjee et al., Appl. Phys. Lett. **74**, 126 (1999).

[10] D. Stauffer and A. Aharony, Introduction to Percolation Theory, 2nd ed., (Taylor and Francis, London, 1991); M. B. Isichenko, Rev. Mod. Phys. **64**, 961(1992).

[11] Z. L. Xiao, E. Y. Andrei, and M. J. Higgins, Phys. Rev. Lett. **83**, 1664 (1999).

[12] Y. Paltiel et al. Phys. Rev. Lett. **85**, 3712 (2000).

[13] Z. L. Xiao, E. Y. Andrei, P. Shuk, and M. Greenblatt, Phys. Rev. B **64**, 094511 (2001).

[14] Z. L. Xiao, E. Y. Andrei, P. Shuk, and M. Greenblatt, Phys. Rev. Lett. **85**, 3265 (2000); U. Yaron et al. Nature (London) **376**, 753 (1995).

[15] C. P. Bean, Rev. Mod. Phys. **36**, 31(1964).

[16] A. I. Larkin and Y. N. Ovchinnikov, J. Low Temp. Phys. **34**, 409 (1979).

[17] L. Burlachkov, Phys. Rev. B **47**, 8056 (1993); Y. Paltiel et al. Phys. Rev. B **58**, R14763 (1998); Z. L. Xiao, E. Y. Andrei, Y. Paltiel, E. Zeldov, P. Shuk, and M. Greenblatt, Phys. Rev. B. **65**, 094511 (2002).

[18] B. I. Halperin, S. Feng, and P. N. Sen, Phys. Rev. Lett. **54**, 2391 (1985); S. Feng, B.I. Halperin, and P.N. Sen, Phys. Rev. B **35**, 197 (1987); B.I. Halperin, Physica D **38**, 179 (1989).

[19] The $I_c$ vs $T$ curve for the annealed lattices follow the SCR line in Figure 1 shown by the open triangles. Thus the onset temperature for the annealed case is 4.7 K.

[20] This is in contrast to the dynamically driven percolation model proposed to describe the scaling of V-I curves at the onset of vortex motion in the liquid state: M. J. Higgins and S. Bhattacharya, Phys. Rev. B. **49**, 10005 (1994); Snarski *et al*., JETP Lett **61**, 119 (1995); J. Watson and D. Fisher, Phys. Rev. B. **55**, 14909 (1997).

FIG. 1.

(a) Time evolution of voltage in response to current pulses corresponding to 3 points on the *V-I* curve at T= 5.035 K and a field of 0.5 T. The narrow voltage spikes at the onset and removal of the pulse are instrumental. (b) *V-I* curves for vortex states prepared by ZFC at temperatures indicated in the legend and measured at $T_0$ = 4.3 K.

FIG. 2.

Dependence of critical currents on measurement speed and method of preparation. Solid circles: FC state obtained with short current pulses; open triangles: FC states obtained with fast current ramps (FCR); open squares: ZFC state obtained with slow current ramps (SCR); solid circles: ZFC obtained with FCR. The inset shows the response of an FC state to short pulses at T = 4.6 K, where the response is ~2 µV for an applied current I = 310 mA, hence $I_c$ (4.6 K) = 310 mA.

FIG. 3.

(a) Temperature dependence of fraction of disordered phase. (b) Temperature dependence of measured and calculated $I_c$. The onset of increase in critical current

(onset of the peak), marked by the arrow, is identified with the percolation transition at $P_c = 0.27$.

FIG. 4.

The dependence of the minimal cross section on $P$ together with the fits to $\alpha_m \propto (P - P_c)^r$ for three different vortex samples. (a) Vortex lattice prepared by ZFC at 0.5 T. (b) Annealed vortex lattice at 0.5 T. (c) ZFC vortex lattice at 0.75 T.

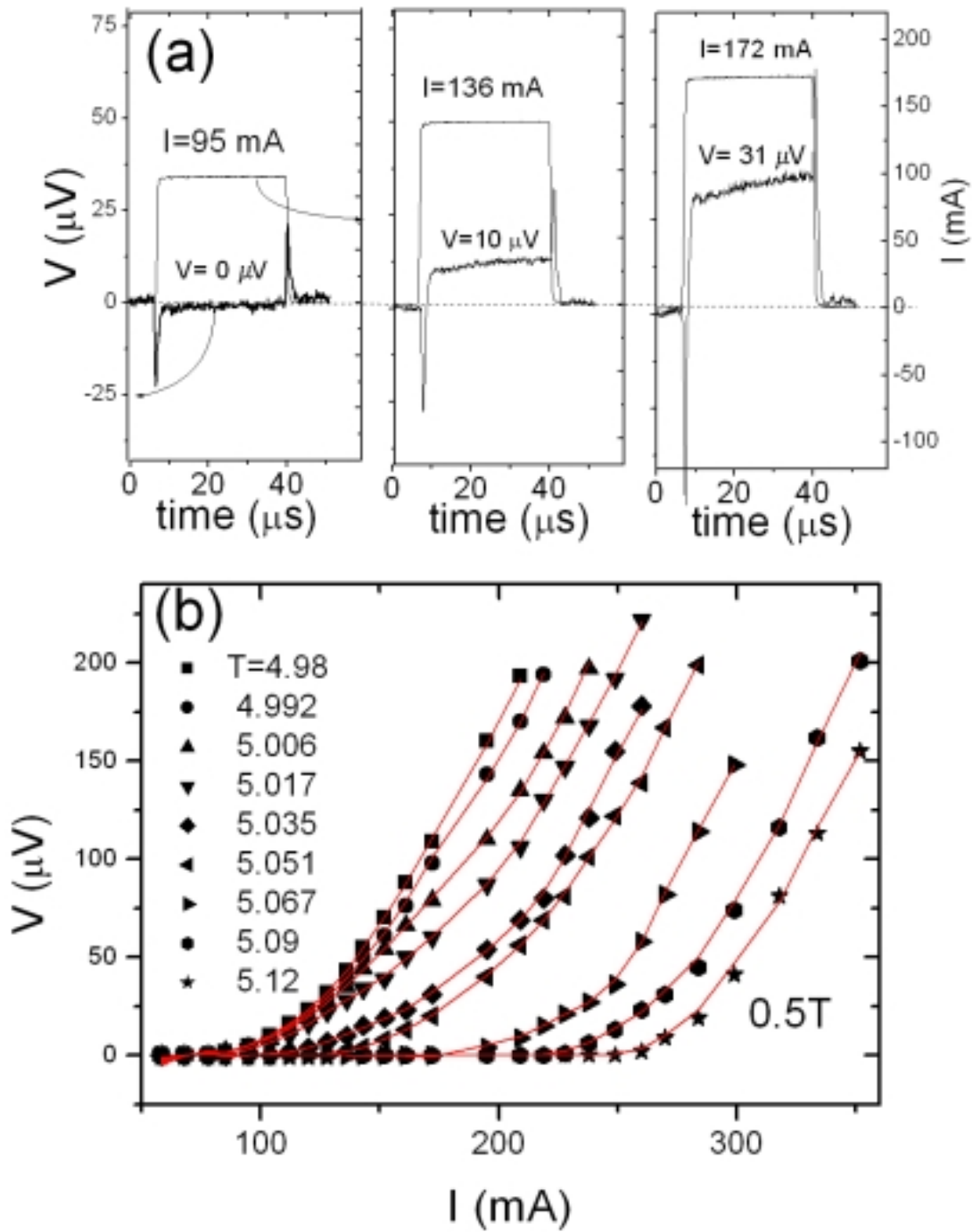

**FIG.1.**

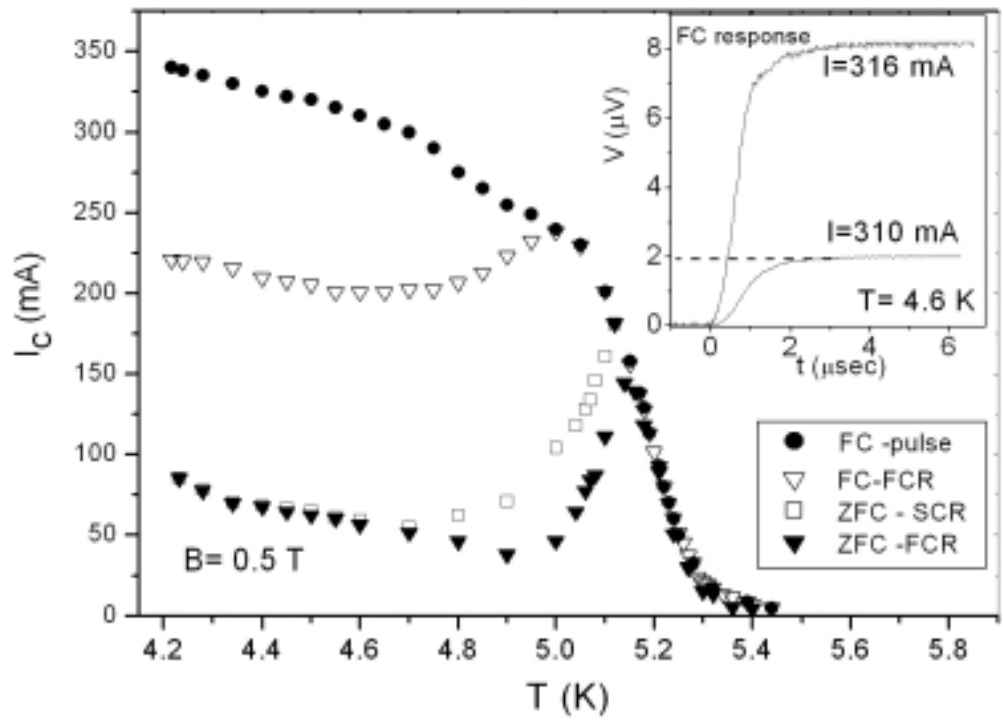

**FIG. 2.**

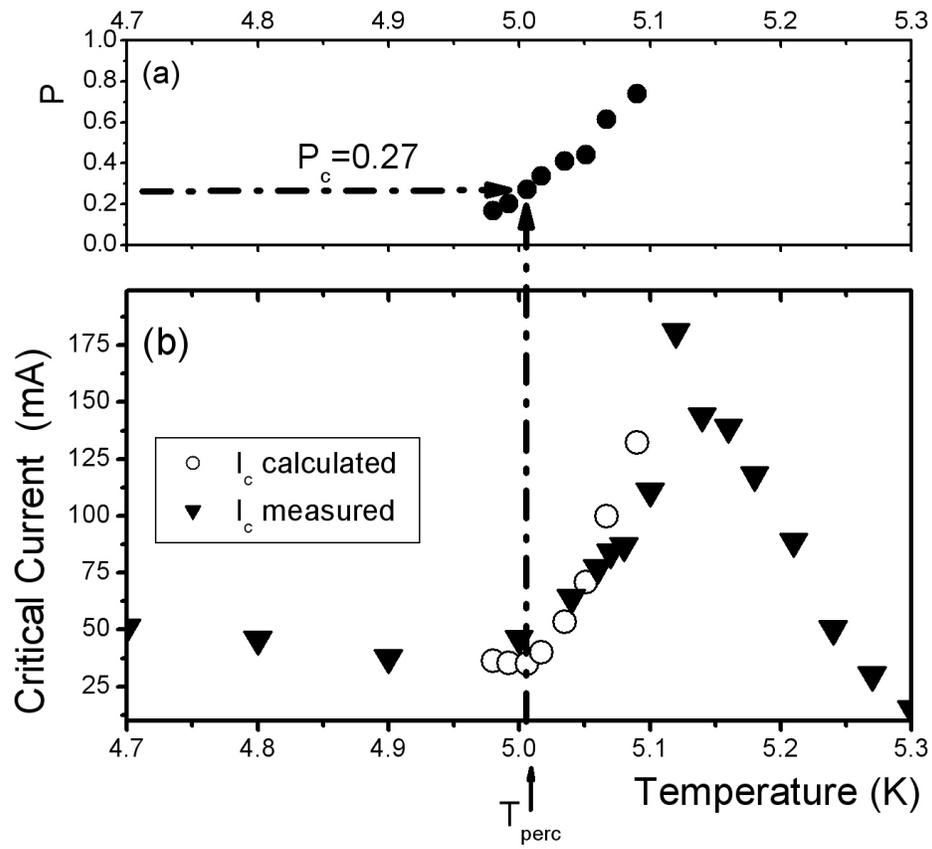

**FIG.3.**

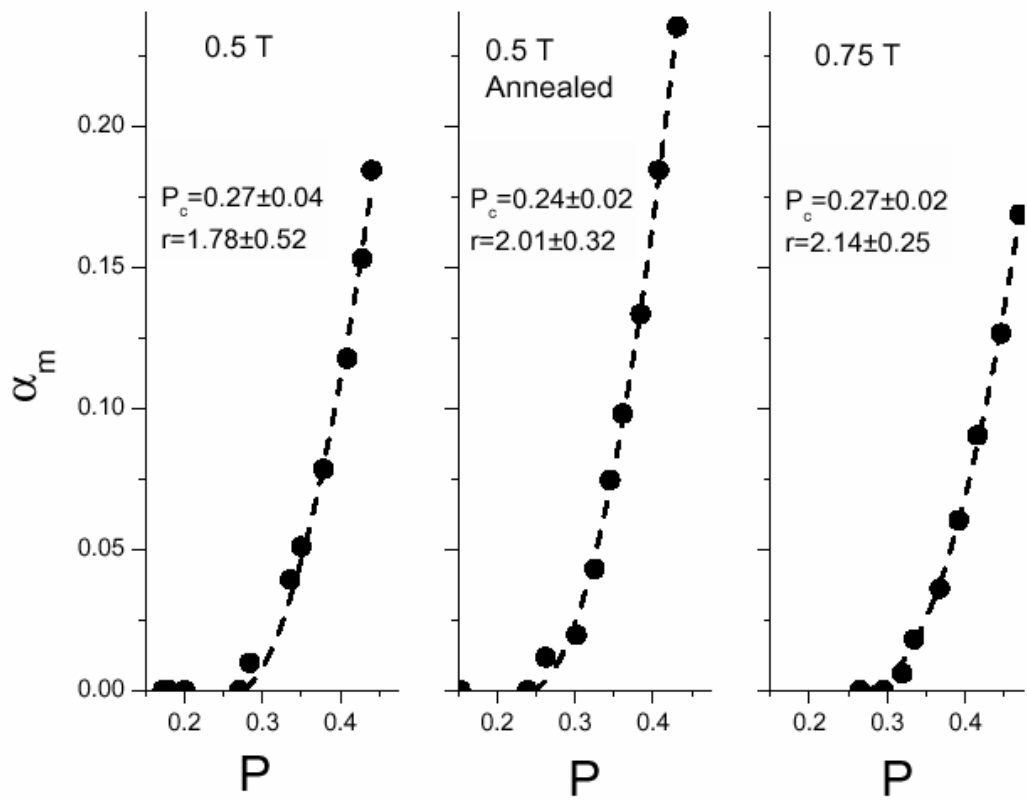

**FIG.4.**